\documentclass[12pt]{article}
\usepackage{a4wide,amsmath,amssymb,graphicx}

\def\ie{i.e.\ }
\def\eg{e.g.\ }
\def\FA{\textsl{FeynArts}}

\def\FC{\textsl{FormCalc}}

\def\LT{\textsl{LoopTools}}
\def\MW{M_W}
\def\MZ{M_Z}
\def\MA{M_{A^0}}
\def\MSUSY{M_{\text{SUSY}}}
\def\sw{s_W}
\def\cw{c_W}
\def\eeWW{e^+e^-\to W^+W^-}
\def\ri{\mathrm{i}}
\def\d{\mathrm{d}}

\def\O{\mathcal{O}}
\def\M{\mathcal{M}}
\def\unity{{\rm 1\mskip-4.25mu l}}
\def\diag{\mathop{\rm diag}}
\def\Re{\mathop{\rm Re}}
\def\GeV{\text{ GeV}}
\def\MeV{\text{ MeV}}
\def\eps{\varepsilon}

\makeatletter
\def\reportno#1{\gdef\@reportno{#1}}
\def\@maketitle{%
  \hfill{\small\begin{tabular}[t]{r}%
    \@reportno
  \end{tabular}\par}%
  \vskip 2em%
  \begin{center}%
    \let \footnote \thanks
    {\large \@title \par}%
    \vskip 1.5em%
    {
      \lineskip .5em%
      \begin{tabular}[t]{c}%
        \@author  
      \end{tabular}\par}%
    \vskip 1em%
    {
     \@date}%
  \end{center}%
  \par
  \vskip 1.5em}
\makeatother

\begin{document}

\reportno{KA--TP--12--2000\\
hep--ph/0007062}

\title{Complete one-loop corrections to $e^+e^-\to W^+W^-$ in the MSSM}

\author{T. Hahn\\
Institut f\"ur Theoretische Physik, Universit\"at Karlsruhe\\
D--76128 Karlsruhe, Germany}

\maketitle

\begin{abstract}
The complete $\O(\alpha)$ corrections including soft-photon bremsstrahlung
to the process $\eeWW$ in the MSSM are calculated for on-shell W bosons.
The relative difference between the MSSM and Standard Model corrections
is generally quite small. The maximum deviation from the Standard Model
within the scanned region of parameter space is $\lesssim 1.5\%$ for
unpolarized and transversally polarized W bosons, and $\lesssim 2.7\%$
for longitudinal W bosons.
\\[1ex]
\end{abstract}


\section{Introduction}

The process $\eeWW$ is already one of the key processes at LEP2, and will
be of similar importance at future linear $e^+e^-$ colliders. Hence it is
not surprising that considerable theoretical effort has gone into the
precise prediction of the cross-section in the Standard Model (SM), both
for on- and off-shell W bosons (\cite{LeV80}, see \cite{BeD94} for a
review).

For a process well accessible both experimentally and theoretically in the
SM, one of the obvious questions to ask is whether it can tell us anything
about physics beyond the SM. Supersymmetric extensions play a special role
because they, like the SM, allow to make precise predictions in terms of a
set of input parameters. Previous calculations in supersymmetric theories
include the complete one-loop corrections in spontaneously broken
supersymmetry \cite{Al94}, sfermion-loop effects in the MSSM
\cite{AlHKSU00,BaDK00}, and also the complete MSSM corrections to the
closely related triple-gauge-boson vertex \cite{ArKM96}.

In this paper the complete one-loop corrections for $\eeWW$ in the MSSM
including real bremsstrahlung in the soft-photon approximation are
presented. The results are rather small: the maximum deviation from the
Standard Model within the scanned region of parameter space is $\lesssim
1.5\%$ for unpolarized and transversally polarized W bosons, and $\lesssim
2.7\%$ for longitudinal W bosons.

The outline of this paper is as follows. In Sect.\ \ref{sect:kin&not} the
kinematics and notation are fixed. Sect.\ \ref{sect:1loop} describes the
details of the one-loop calculation. The results of the calculation and
the scan over the MSSM parameter space are presented in Sect.\
\ref{sect:results}. Sect.\ \ref{sect:conclus} finally gives the
conclusions.

\section{Kinematics and notation}
\label{sect:kin&not}

The reaction studied here is
\begin{equation}
e^+(k_1,\lambda_1) + e^-(k_2,\lambda_2) \to
W^+(k_3,\lambda_3) + W^-(k_4,\lambda_4)\,,
\end{equation}
where $k_i$ and $\lambda_i$ represent the momenta and helicities of the
external particles, respectively.

The incoming particles travel along the $z$ axis and are scattered into
the $x$--$z$ plane. Neglecting the electron mass, the explicit
representations of the momenta and the polarization vectors in the
centre-of-mass system are
\begin{equation}
\label{eq:vecs}
\begin{aligned}
k_1^\mu &= E \left(1,\, 0,\, 0,\, -\beta\right), &
\eps_3^\mu(0) &=
  \left(-p,\, E\sin\theta,\, 0,\, E\cos\theta\right)/\MW\,, \\
k_2^\mu &= E \left(1,\, 0,\, 0,\,  \beta\right), &
\eps_3^\mu(\pm) &=
  \left(0,\, -\cos\theta,\, \pm\ri,\, \sin\theta\right)/\sqrt 2\,, \\
k_3^\mu &= \left(E,\, -p\sin\theta,\, 0,\, -p\cos\theta\right), \qquad &
\eps_4^\mu(0) &=
  \left(-p,\, -E\sin\theta,\, 0,\, -E\cos\theta\right)/\MW\,, \\
k_4^\mu &= \left(E,\,  p\sin\theta,\, 0,\,  p\cos\theta\right), &
\eps_4^\mu(\pm) &=
  \left(0,\, \cos\theta,\, \pm\ri,\, -\sin\theta\right)/\sqrt 2\,,
\end{aligned}
\end{equation}
where $\beta$ is the velocity of the electrons, $p = \sqrt{E^2 - \MW^2}$
is the momentum of the W bosons, $E = \sqrt s/2$ is the beam energy, and
$\theta$ is the scattering angle.

The polarized differential cross-section is obtained from the helicity
amplitudes $\M$ as
\begin{equation}
\label{eq:diffWQ}
\left(\frac{\d\sigma}{\d\Omega}\right)_
    {\lambda_1\lambda_2\lambda_3\lambda_4}
= \frac 1{64\pi^2 s}
  \left|\M_{\lambda_1\lambda_2\lambda_3\lambda_4}\right|^2.
\end{equation}
In the following, only unpolarized electrons are considered, since that is
the situation at LEP2. In this case the cross-section has to be averaged
over the initial helicities according to
\begin{equation}
\left(\frac{\d\sigma}{\d\Omega}\right)_{\text{UU}\lambda_3\lambda_4} =
\frac 14 \sum_{\lambda_1,\lambda_2 = \pm 1}
\left(\frac{\d\sigma}{\d\Omega}\right)_
  {\lambda_1\lambda_2\lambda_3\lambda_4}.
\end{equation}
Polarization combinations are given by a sequence of four letters, \eg
UULT, where L, T, and U denote longitudinal ($\lambda = 0$), transverse
($\lambda = \pm$), and unpolarized particles.

\section{One-loop corrections}
\label{sect:1loop}

\begin{figure}
\begin{center}
\includegraphics{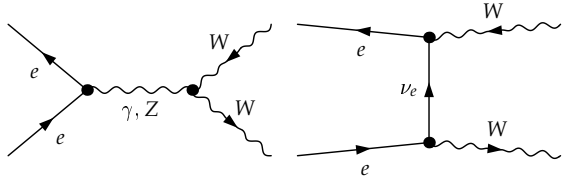}
\end{center}
\caption{\label{fig:born}The tree-level diagrams.}
\end{figure}

Since the coupling of the electron to any of the Higgs particles in the
SM or the MSSM is suppressed by a factor $m_e/\MW$, the
Higgs-exchange diagrams can safely be neglected. The tree-level diagrams
are then the same for the SM and the MSSM, \ie $\gamma$ and
$Z$ exchange in the $s$-channel and neutrino exchange in the $t$-channel
(see Fig.\ \ref{fig:born}). The tree-level results shall not be discussed
here as this has been done in detail elsewhere (explicit formulas can be
found \eg in \cite{BeD94}).

\subsection{Calculational framework}

The calculation of the $\O(\alpha)$ radiative corrections has been
performed in 't~Hooft--Feynman gauge. Ultraviolet (UV) divergences were
treated within dimensional regularization. For the renormalization the
on-shell scheme \cite{RoT73} was used, following the formulation worked
out in \cite{De93}. In the absence of SUSY particles at tree level, only
those counter-terms appear that are already present in the SM. The
only change is that now the self-energies from which the renormalization
constants are derived have to be calculated in the MSSM.

To $\O(\alpha)$, the squared matrix element is given by
\begin{equation}
|\M|^2 = |\M_{\text{Born}}|^2 (1 + \delta_{\text{soft}}) +
2\Re\left(\M_{\text{Born}}^*\M_{\text{1-loop}}\right)\,,
\end{equation}
where $\M_{\text{Born}}$ and $\M_{\text{1-loop}}$ denote the sum of the
contributing tree-level and one-loop Feynman diagrams, respectively, and
$\delta_{\text{soft}}$ is the QED correction factor from real
bremsstrahlung in the soft-photon approximation.

The Feynman diagrams were generated with \FA\ \cite{KuBD91} which in its
current version uses algorithms that can deal with supersymmetric theories
\cite{Ha00}. The resulting amplitudes were algebraically simplified using
\FC\ \cite{HaP98} and then converted to a Fortran program. The \LT\
package \cite{HaP98, vOV90} was used to evaluate the one-loop scalar and
tensor integrals. For a single point in phase and parameter space the
Fortran program runs about 3.6 ms in the SM and 102 ms in the MSSM.  
Scans over parameter space in the MSSM typically take several hours.

Apart from obvious checks such as UV- and IR-finiteness, the SM results of
\cite{BeD94} were fully confirmed and the effects of sfermion-loops in
\cite{AlHKSU00} could be reproduced for various scenarios of sfermion
masses to within a very small constant shift ($\sim .5\%$) which is an
effect of the different renormalization schemes used and hence of higher
order.

With the presence of external gauge bosons one might expect sizable gauge
cancellations and hence instabilities in the numerical code. To safeguard
against such problems, the SM values were computed once in the
conventional way and once in the background-field method \cite{DeDW95},
where gauge cancellations should be greatly reduced, and an agreement to 9
digits was found. Even though the same comparison cannot easily be done
for the MSSM (a corresponding model file is currently not available), a
similar stability is expected since the gauge structure of the MSSM is
essentially the same as in the SM. As an alternative measure of stability,
one can compare the individual contributions from self-energies, vertices,
and boxes with their sum. To show that the cancellations are harmless
numerically, the following table lists the individual contributions (each
of them renormalized) that make up the amplitude at a particular point in
parameter space for longitudinal W bosons, for which the largest
cancellations can be expected.
\begin{equation}
\label{eq:gaugecanc}
\left.\frac 1{64\pi^2 s} 2\Re\M_{\text{Born}}^*\M_X
\right|_{\begin{subarray}{l}
\sqrt s = 500\,\text{GeV} \\
\theta = 90^\circ
\end{subarray}}
= \left\{
\begin{array}{lr}
X = \text{self-energies:} & -.02535074~\text{pb} \\
X = \text{vertices:}      & -.03675248~\text{pb} \\
X = \text{boxes:}         &  .06422023~\text{pb} \\
\hline
X = \text{sum:}           &  .00211700~\text{pb}
\end{array}\right.
\end{equation}

\subsection{QED corrections}
\label{sect:qed}

In addition to the virtual diagrams, real photon emission from the
external legs has to be taken into account to cancel the infrared (IR)
divergences which arise from the exchange of massless photons. The IR
divergences are regularized by an infinitesimal photon mass $\lambda$.
In the soft-photon limit the cross-section for real photon emission is
proportional to the Born cross-section,
\begin{equation}
\left(\frac{\d\sigma}{\d\Omega}\right)_{\text{soft}} =
\left(\frac{\d\sigma}{\d\Omega}\right)_{\text{Born}}
\delta_{\text{soft}}\,,
\end{equation}
with the soft-photon factor $\delta_{\text{soft}}$ given by
\begin{equation}
\label{eq:deltasoft}
\delta_{\text{soft}} = -\frac{e^2}{(2\pi)^3}
\int\limits_{k_0\leqslant\Delta E}\frac{\d^3 k}{2 k_0}
\left.
  \sum_{i,j = 1}^4\frac{\pm Q_i Q_j (k_i k_j)}{(k_i k) (k_j k)}
\right|_{k_0 = \sqrt{k^2 + \lambda^2}}\,.
\end{equation}
Here $\Delta E$ is the maximum energy of the emitted photons, $Q_i$ is the
charge of the $i$th external particle, and the sign in front of the
product $Q_i Q_j$ is $+$ if particles $i$ and $j$ are both either incoming
or outgoing, and $-$ otherwise. The basic integrals needed for the
evaluation of \eqref{eq:deltasoft} have been worked out \eg in
\cite{tHV79}.

The absolute magnitude of the $\O(\alpha)$ corrections is largely
determined by the QED contributions and depends strongly on the
photon-energy cutoff $\Delta E$ through logarithms of the form
$\log\Delta E/E$. Without an experimentally motivated choice of $\Delta
E$, the overall size of the one-loop corrections can be shifted more or
less at will using different values of $\Delta E$. Since the intention
here is only to show the differences between the SM and the MSSM
calculations, the concrete value of $\Delta E$ is rather unimportant.
In this calculation a soft-photon cutoff energy of
$\Delta E = 0.05\sqrt s$ has been used as in \cite{BeD94}.

The cancellation of the IR divergences has been checked numerically by
varying the photon-mass $\lambda$ on which the result must not depend. A
variation of $\lambda$ from 1 to $10^{10}$ GeV leaves the first 10 digits
of the cross-section invariant.

While soft-photon radiation is sufficient to cancel IR divergences, it is
an approximation, valid only if $\Delta E$ is small compared to all
relevant energy scales. If this is not the case, hard-photon radiation
must be taken into account, too. However, the hard-photon corrections at
$\O(\alpha)$ are exactly the same as in the SM, therefore they have been
omitted in the present calculation.

\subsection{Inventory of one-loop diagrams}

This section lists the contributing one-loop diagrams. The diagrams can
be separated into self-energy contributions, vertex corrections, and box
corrections. Note that diagrams which are already present in the SM are
omitted in the figures.

\begin{figure}
\begin{center}
\includegraphics{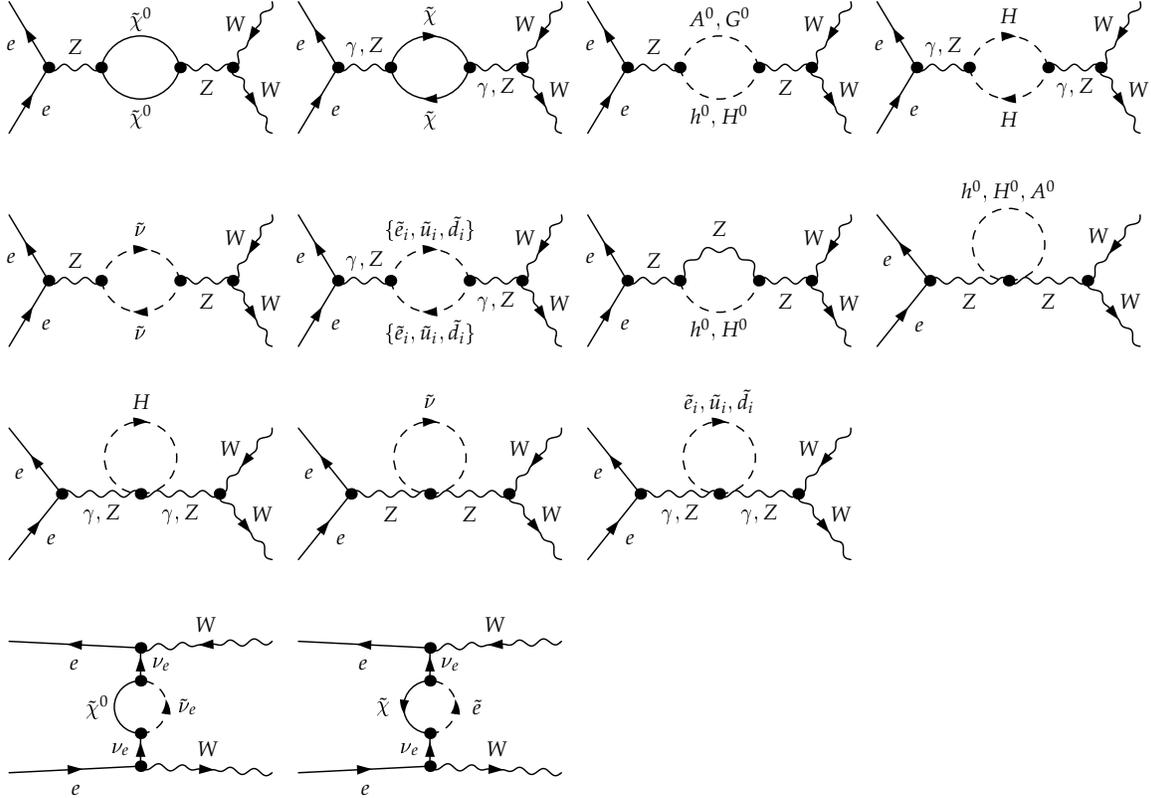}
\end{center}
\caption{\label{fig:selfdiags}The MSSM self-energy corrections.
Braces indicate that there is one diagram for the first members of all
braced lists, one for the second members, etc. A sfermion with index $i$
accounts for six particles, \eg $\tilde e_i = \{\tilde e^1, \tilde e^2,
\tilde\mu^1, \tilde\mu^2, \tilde\tau^1, \tilde\tau^2\}$.}
\end{figure}

The self-energy contributions fall into two categories, corrections to the
$\gamma$ and $Z$ propagator in the $s$-channel, and corrections to the
$\nu$ propagator in the $t$-channel. These diagrams are shown in Fig.\
\ref{fig:selfdiags}.

\begin{figure}
\begin{center}
\includegraphics{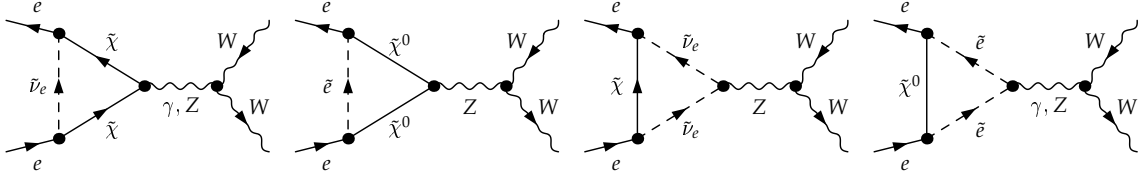}
\end{center}
\caption{\label{fig:vertsidiags}The MSSM contributions to the
initial-state vertex in the $s$-channel.}
\end{figure}

\begin{figure}
\begin{center}
\includegraphics{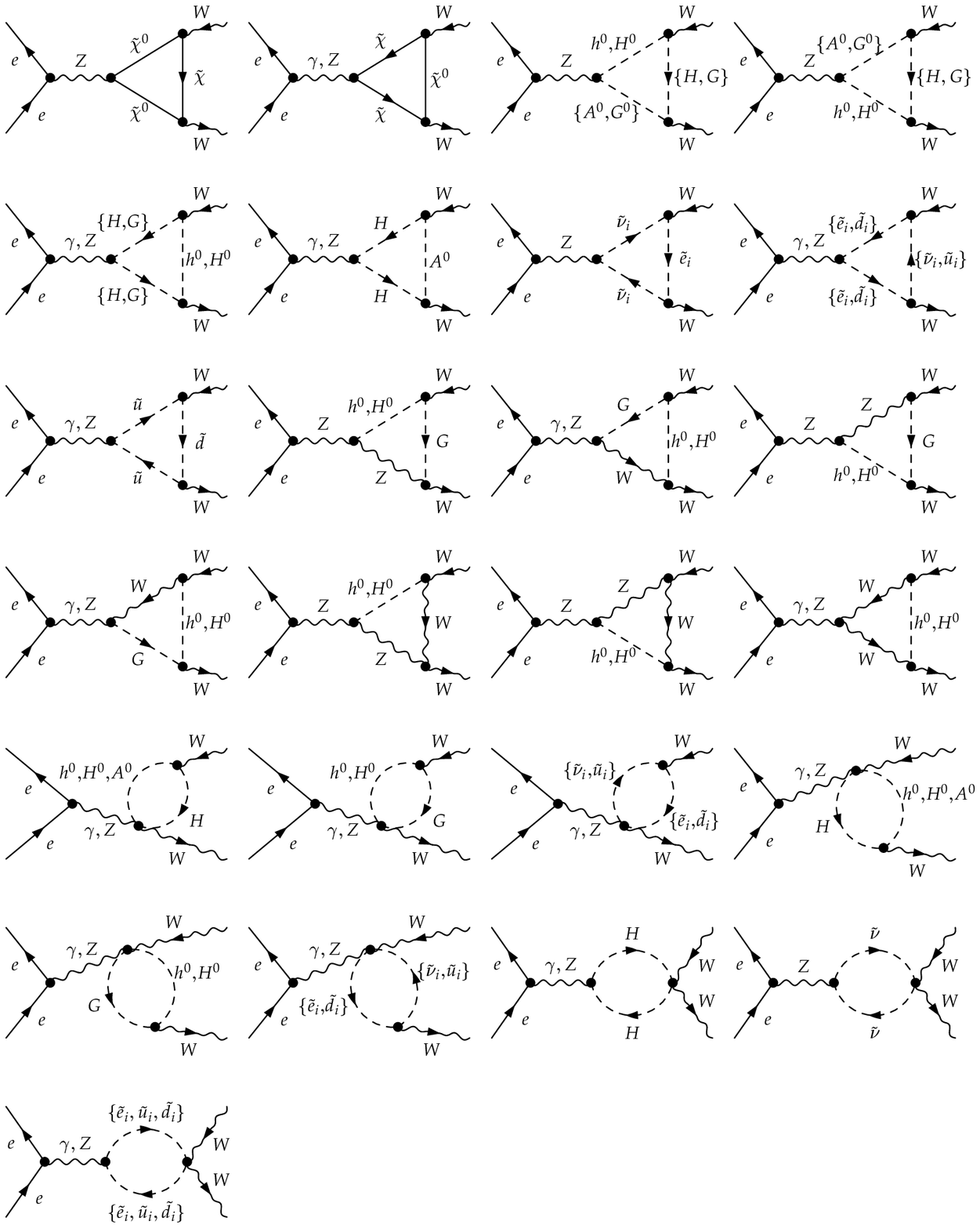}
\end{center}
\caption{\label{fig:vertsfdiags}The MSSM contributions to the $s$-channel 
final-state vertex. Braces indicate that there is one diagram for the
first members of all braced lists, one for the second members, etc. A
sfermion with index $i$ accounts for six particles, \eg $\tilde e_i =
\{\tilde e^1, \tilde e^2, \tilde\mu^1, \tilde\mu^2, \tilde\tau^1,
\tilde\tau^2\}$.}
\end{figure}

\begin{figure}
\begin{center}
\includegraphics{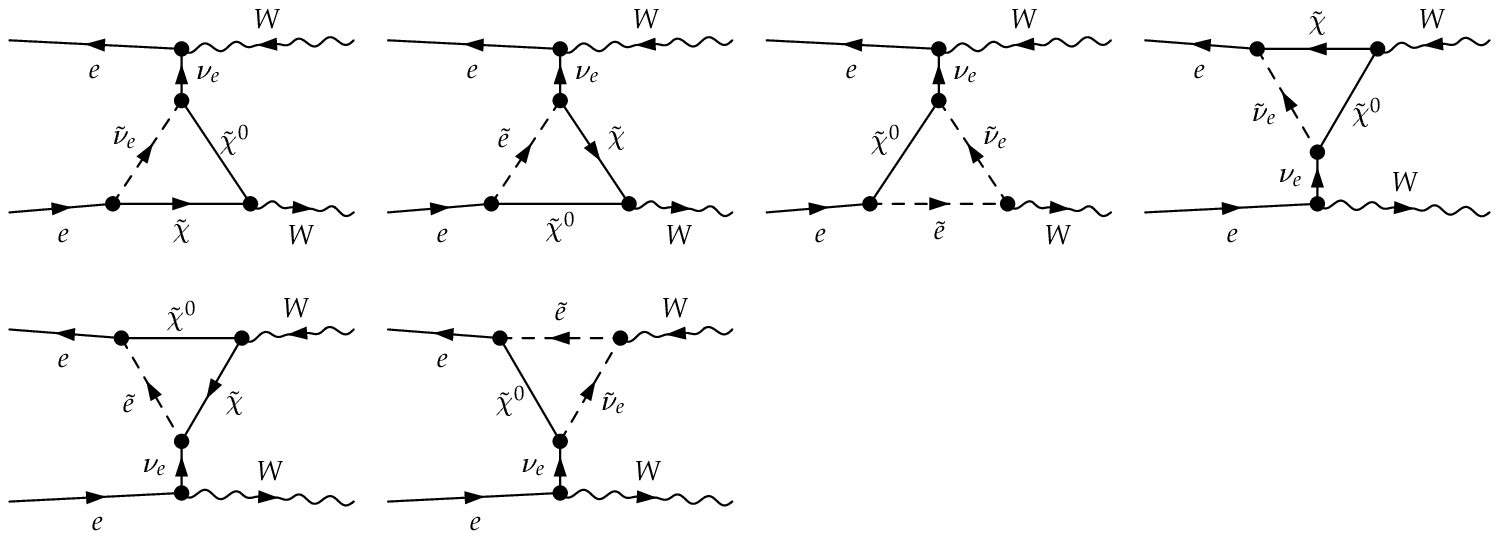}
\end{center}
\caption{\label{fig:verttdiags}The MSSM contributions to the $t$-channel
vertices.}
\end{figure}

The vertex diagrams can be grouped into corrections to the initial- and
final-state vertex in the $s$-channel (Figs.\ \ref{fig:vertsidiags} and
\ref{fig:vertsfdiags}), and corrections to the $t$-channel vertices (Fig.\
\ref{fig:verttdiags}).

\begin{figure}
\begin{center}
\includegraphics{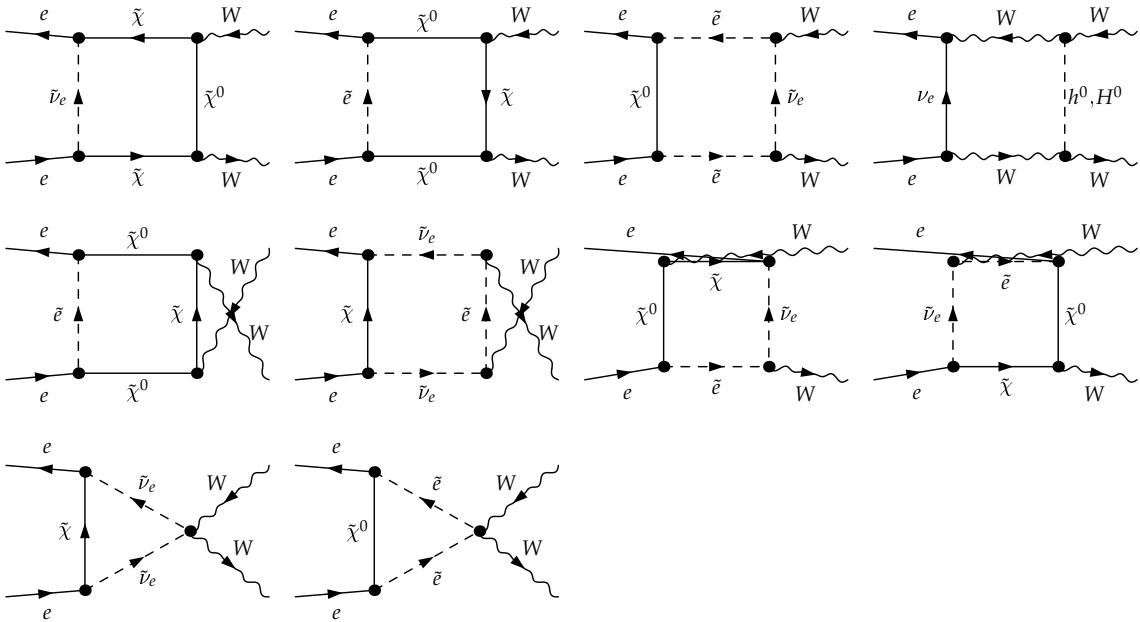}
\end{center}
\caption{\label{fig:boxdiags}The MSSM box diagrams.}
\end{figure}

Finally, the box diagrams are displayed in Fig. \ref{fig:boxdiags}.

\section{Numerical results}
\label{sect:results}

\subsection{Input parameters}

\subsubsection{Standard Model parameters}

For the SM parameters the following numerical values are used:
\begin{equation} 
\begin{aligned}
\alpha^{-1} &= 137.0359895,\quad &
\MZ &= 91.1867\GeV,\quad &
\MW &= 80.39\GeV, \\[.5ex]
m_e &= 0.51099907\MeV,\quad &
m_u &= 53.8\MeV,\quad &
m_d &= 53.8\MeV,\quad \\
m_\mu &= 105.658389\MeV, &
m_c &= 1.50\GeV, &
m_s &= 150\MeV, \\
m_\tau &= 1777\MeV, &
m_t &= 174\GeV, &
m_b &= 4.7\GeV.
\end{aligned} 
\end{equation}
The masses of the up and down quarks are effective parameters which are
adjusted such that the five-flavour hadronic contribution to
$\Delta\alpha$ is 0.02778 \cite{Je99}, \ie
\begin{equation*}
\Delta\alpha_{\text{had}}^{(5)}(s = \MZ^2)
= \frac{\alpha}{\pi}\sum_{f = u,c,d,s,b}
  q_f^2 \Bigl(\log\frac{\MZ^2}{m_f^2} - \frac 53\Bigr)
\overset{!}{=} 0.02778\,.
\end{equation*}

\subsubsection{MSSM parameters}

\paragraph{Higgs sector}

The neutral Higgs sector is fixed by choosing a value for $\tan\beta$ and
for the mass $\MA$ of the CP-odd neutral Higgs boson $A^0$. For the other
neutral-Higgs masses, which receive significant radiative corrections
\cite{HaH91}, the two-loop approximation formula of \cite{HeHW99} is used.

There are only small radiative corrections for the charged Higgs masses
and the following equation holds for $\MA\sim\O(\MW)$ \cite{Diaz92}.
\begin{multline}
\label{mHplus}
M_{H^\pm}^2 = \MA^2 + \MW^2
  + \frac{5\alpha\MW^2}{2\pi\cw^2}\ln\frac{\MSUSY}{\MW} \\
  + \frac{3\alpha}{4\pi\sw^2\MW^2}\left[
      \frac{2 m_{t,r}^2 m_b^2}{\sin^2\!\beta\cos^2\!\beta}
      - \MW^2 \left(\frac{m_{t,r}^2}{\sin^2\!\beta}
                    + \frac{m_b^2}{\cos^2\!\beta}\right)
      + \frac 23 \MW^4
     \right] \ln\frac{\MSUSY}{m_{t,r}}\,,
\end{multline}
where $\MSUSY$ is a universal soft-SUSY-breaking mass introduced in the
next paragraph and $m_{t,r} = m_t (1 + 4\alpha_s(m_t)/3\pi)^{-1}$ is the
running top mass.

\paragraph{Sfermions}

For simplicity, all soft-SUSY-breaking parameters are assumed equal and
mixing between sfermion generations is neglected, so that
\begin{equation}
\label{Msusy}
\begin{gathered}
M_{\tilde Q}^2 = M_{\tilde U}^2 = M_{\tilde D}^2 = 
M_{\tilde L}^2 = M_{\tilde E}^2 = \MSUSY^2\unity\,, \\
A_U = A_u\unity\,, \qquad
A_L = A_D = A_d\unity\,.
\end{gathered}
\end{equation}
Then, the sfermion mass matrix is given by \cite{HaK85,GuH86}
\begin{gather}
\begin{pmatrix}
\MSUSY^2 + M_Z^2 \cos(2\beta) (I_3^f - Q_f\sw^2) + m_f^2 & 
	m_f (A_{\{u,d\}} - \mu\{\cot\beta,\tan\beta\}) \\
m_f (A_{\{u,d\}} - \mu\{\cot\beta,\tan\beta\}) &
	\MSUSY^2 + M_Z^2 \cos(2\beta) Q_f\sw^2 + m_f^2
\end{pmatrix}
\end{gather}
where the elements in braces apply to $I_3^f = +\frac 12$ and $-\frac 12$,
respectively.

\paragraph{Charginos and Neutralinos}

The chargino mass matrix \cite{HaK85,GuH86}
\begin{equation}
X = \begin{pmatrix}
M_2 & \sqrt 2 M_W\sin\beta \\
\sqrt 2 M_W\cos\beta & \mu
\end{pmatrix}
\end{equation}
and the neutralino mass matrix \cite{HaK85,GuH86}
\begin{equation}
\label{Y}
Y = \begin{pmatrix}
M_1 & 0 & -M_Z\sw\cos\beta & M_Z\sw\sin\beta \\
0 & M_2 & M_Z\cw\cos\beta & -M_Z\cw\sin\beta \\
-M_Z\sw\cos\beta & M_Z\cw\cos\beta & 0 & -\mu \\
M_Z\sw\sin\beta & -M_Z\cw\sin\beta & -\mu & 0
\end{pmatrix}
\end{equation}
are diagonalized with unitary matrices $U$, $V$, and $N$ such that
\begin{equation}
\begin{gathered}
U^* X V^{-1} = \diag(m_{\tilde\chi_1}, m_{\tilde\chi_2})\,, \\
N^* Y N^{-1} = \diag(m_{\tilde\chi^0_1}, \ldots, m_{\tilde\chi^0_4})\,.
\end{gathered}
\end{equation}
This diagonalization is done numerically using the subroutines of the
LAPACK library \cite{lapack}. The $U(1)$ gaugino-mass parameter $M_1$
which appears as a further input parameter in $Y$ is fixed, as usual, by
the SUSY-GUT relation
\begin{equation}
\label{M-GUT}
M_1 = \frac 53 \frac{\sw^2}{\cw^2} M_2\,.
\end{equation}

\subsubsection{Parameter scan}

The remaining input parameters are then scanned over the following
regions:
\begin{equation}
\label{eq:scanregions}
\begin{array}{c|c}
\text{SM} & \text{MSSM} \\ \hline
M_H = 100\dots 300 \GeV &
\begin{aligned}
\tan\beta &= 1.5, 5, 50 \\
\MA &= 100\dots 1000\GeV \\
\mu &= -1000\dots 1000\GeV \\
M_2 &= 100\dots 1000\GeV \\
\MSUSY &= 100\dots 1000\GeV \\
A_u = A_d &= \MSUSY, 2\MSUSY
\end{aligned}
\end{array}
\end{equation}
When the particle masses are calculated from these input parameters, the
Fortran program checks whether they are consistent with the current
exclusion limits and automatically omits already-excluded points in
parameter space from the scan. The following bounds are used:
\begin{equation}
\begin{aligned}
M_{\tilde t} &\geqslant 80\GeV~\cite{squarkbounds}\,, \qquad &
M_{h^0} &\geqslant 85\GeV~\cite{h0bounds}\,, \\
M_{\tilde b} &\geqslant 70\GeV~\cite{squarkbounds}\,, &
m_{\tilde\chi^0} &\geqslant 30\GeV~\cite{chibounds}\,, \\
M_{\tilde q\neq \tilde b, \tilde t}
  &\geqslant 150\GeV~\cite{squarkbounds}\,, &
m_{\tilde\chi} &\geqslant 90\GeV~\cite{chibounds}\,. \\
M_{\tilde\ell} &\geqslant 70\GeV~\cite{sleptonbounds}\,,
\end{aligned}
\end{equation}

\subsection{Results}

\begin{figure}
\begin{center}
\includegraphics{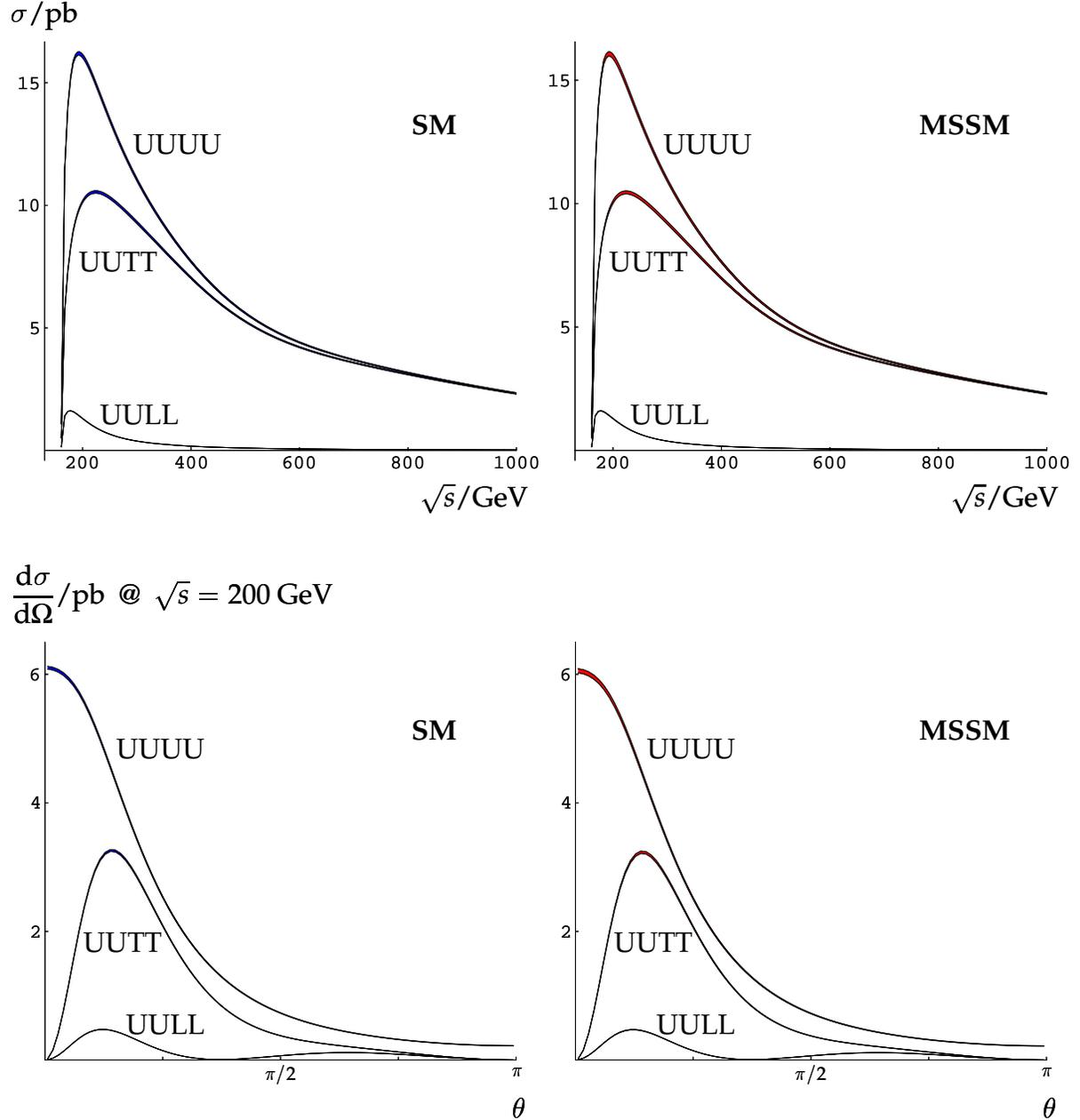}
\end{center}
\caption{\label{fig:absscan}The total cross-section and the differential
cross-section at $\sqrt s = 200$ GeV for the SM and the MSSM. The
thickness of the curves give the range of allowed values within the
parameter scan \eqref{eq:scanregions}. The polarizations are denoted by:
UUUU (all particles unpolarized), UUTT (transverse W bosons), and UULL
(longitudinal W bosons).}
\end{figure}

In Fig.\ \ref{fig:absscan} the absolute values of the total and
differential cross-section in the SM and MSSM is shown for three different
polarizations: UUUU (all particles unpolarized), UUTT (transverse W
bosons), and UULL (longitudinal W bosons). The thickness of the curves
reflects the range of the allowed values within the parameter scan
\eqref{eq:scanregions}.

\begin{figure}
\begin{center}
\includegraphics{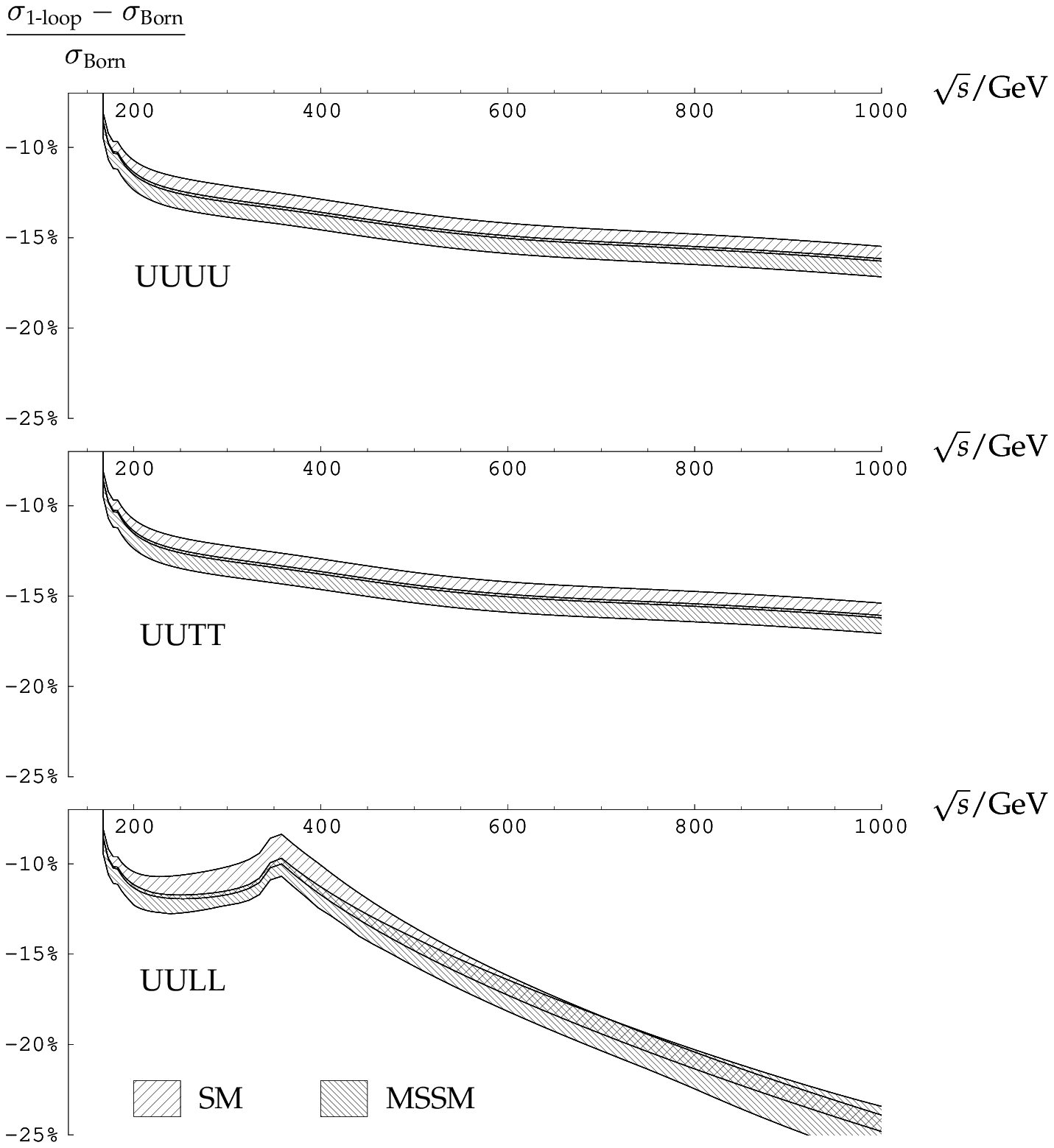}
\end{center}
\caption{\label{fig:relscan}The relative difference between the one-loop
corrected and the Born cross-section. The cross-hatched bands indicate the
minimum and maximum reached within the scan over the parameter space
\eqref{eq:scanregions}. The polarizations are denoted by: UUUU (all
particles unpolarized), UUTT (transverse W bosons), and UULL (longitudinal
W bosons).}
\end{figure}

Fig.\ \ref{fig:relscan} shows the relative corrections in the SM and MSSM
with respect to the Born cross-section. The overall magnitude of the
corrections is largely determined by the well-understood QED logarithms,
as discussed in Sect.\ \ref{sect:qed}. The bands show the variation of the
cross-section within the parameter scan \eqref{eq:scanregions}. The bands
for the SM and MSSM overlap slightly, the MSSM band being the lower one.
The two bands are of comparable width of the order of a few percent for
all polarizations. The dominating polarizations over the entire energy
range are UU$\pm$$\mp$, and this explains why the plots for the UUUU and
UUTT polarizations in Fig.\ \ref{fig:relscan} are quite similar. On the
other hand it is not possible to trace the origins of the radiative
corrections by disentangling the self-energy, vertex, and box
contributions since in the presence of external gauge bosons there are
delicate gauge cancellations.

\begin{figure}
\begin{center}
\includegraphics{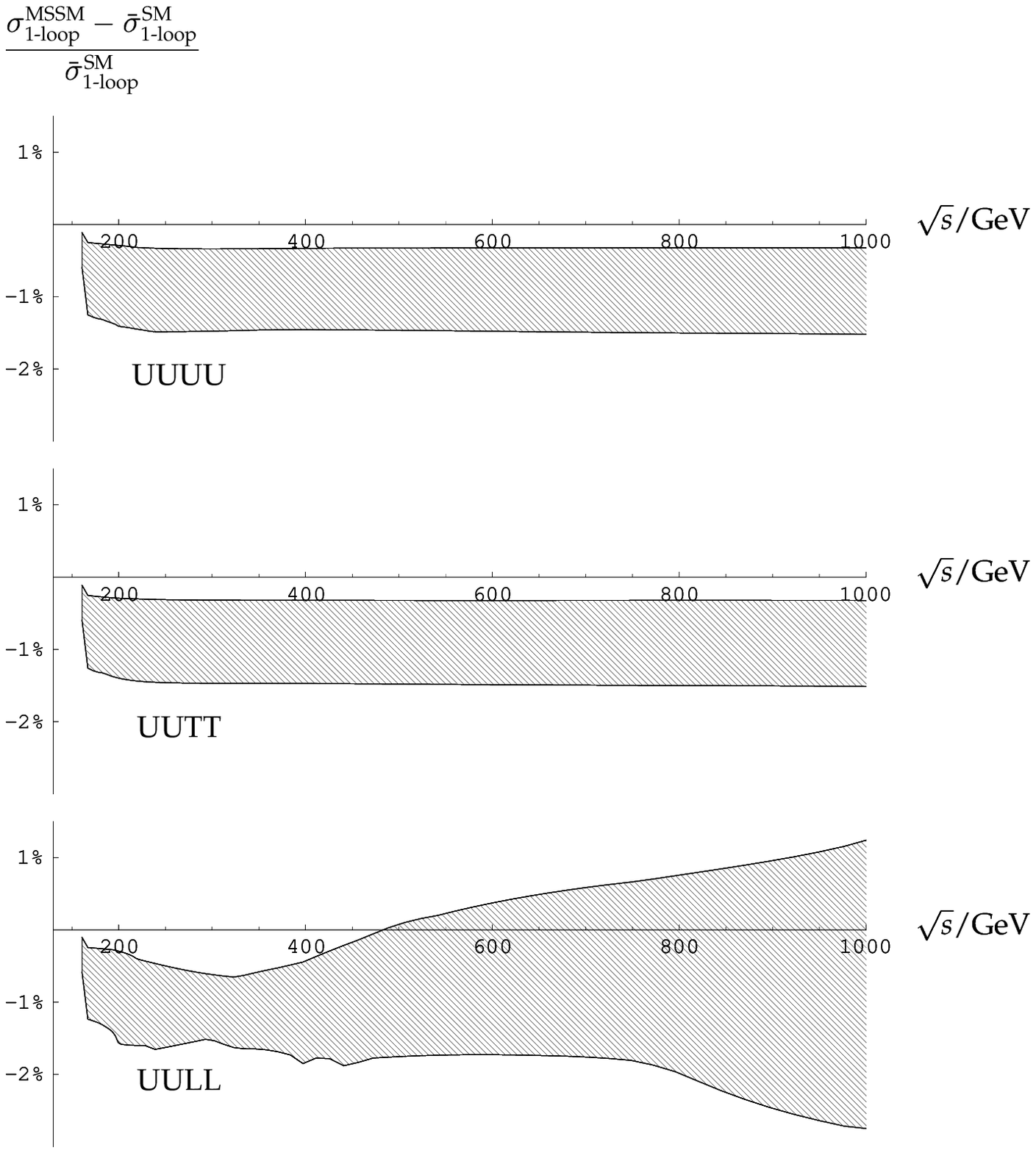}
\end{center}
\caption{\label{fig:smmssm}The relative difference between the SM central
value $\bar\sigma_{\text{1-loop}}^{\text{SM}}$ and the MSSM bands of
Fig.\ \ref{fig:relscan}. The polarizations are denoted by: UUUU (all
particles unpolarized), UUTT (transverse W bosons), and UULL (longitudinal
W bosons).}
\end{figure}

The relative deviation between the SM central value and the MSSM band is
plotted in Fig.\ \ref{fig:smmssm}. While the largest corrections ($\sim
-2.7\%$ at 1 TeV) are seen for purely longitudinal W bosons, one has to
keep in mind that the cross-section for longitudinally polarized W bosons
is much smaller than for the transverse polarizations. In the transverse
polarizations and in the unpolarized case the maximum deviation between
the SM and the MSSM is roughly 1.5\%. The maximum deviation is reached in
all polarizations for light SUSY particles, for example in the unpolarized
case (UUUU) the lightest higgs, stop, chargino, and neutralino masses at
the point of maximum deviation are $M_{h^0} = 91\GeV$, $M_{\tilde t} =
165\GeV$, $m_{\tilde\chi} = 117\GeV$, and $m_{\tilde\chi^0} = 52\GeV$,
respectively.

\begin{figure}
\begin{center}
\includegraphics{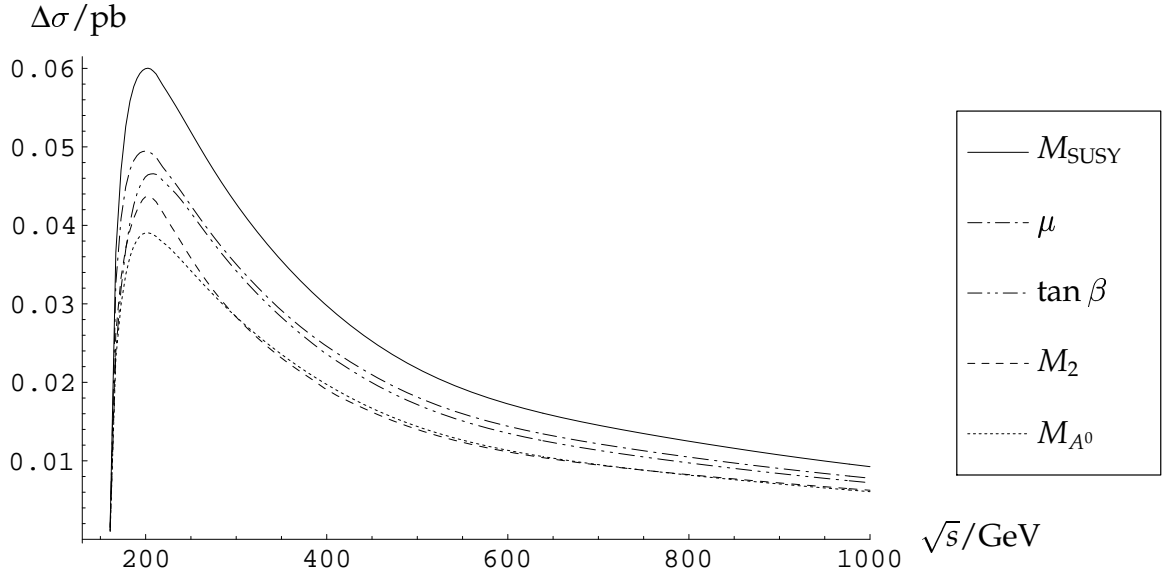}
\end{center}
\caption{\label{fig:variation}The variation of the MSSM cross-section for
unpolarized particles with the scanned MSSM parameters. The variation of a
parameter $X$ is computed as the difference between maximum and minimum of
the cross-section for fixed $X$, averaged over all values of $X$.}
\end{figure}

The variation of the cross-section with the scanned MSSM parameters is
shown in Fig.\ \ref{fig:variation}. The largest variation is connected
with the soft-SUSY-breaking mass $\MSUSY$ and confirms the idea of the
previous calculations \cite{AlHKSU00, BaDK00} that important contributions
come from the sfermion sector. Nevertheless, the contributions from the
other sectors are similar in size and cannot be neglected.

\begin{figure}
\begin{center}
\includegraphics{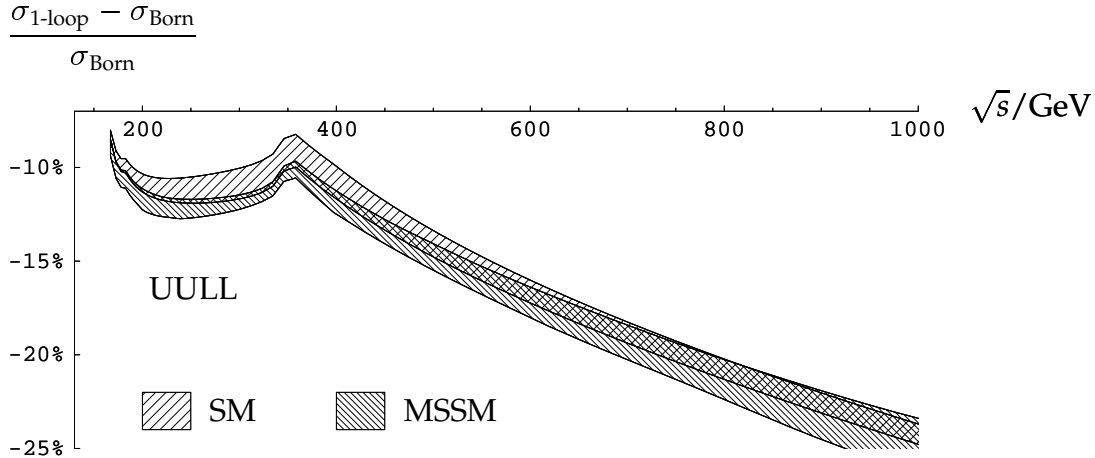}
\end{center}
\caption{\label{fig:relscanGF}The same as the lower diagram in Fig.\ 
\ref{fig:relscan} (\ie the relative one-loop corrections for longitudinal 
W bosons), but calculated from the input parameter set $\{\alpha, G_F,
\MZ\}$ instead of $\{\alpha, \MW, \MZ\}$. The cross-hatched bands indicate
the minimum and maximum reached within the scan over the parameter space
\eqref{eq:scanregions}.}
\end{figure}

The calculation was checked for a possible scheme dependence by repeating
it using the input values $\{\alpha, G_F, \MZ\}$ instead of $\{\alpha,
\MW, \MZ\}$. This is not entirely straightforward since $\MW$ enters not
only in loop corrections, but more directly in the kinematics of the
process. For instance, the threshold of $2\MW$ is different in the SM and 
MSSM so that it is not admissible to plot the ratio as in Fig.\ 
\ref{fig:smmssm}. To compute $\MW(\alpha, G_F, \MZ)$, the fit formula of
\cite{FrHHWW01}, approximating the two-loop result, together with the MSSM
contributions to $\Delta\rho$ \cite{DjGHHJW97} have been used. The results
are shown for longitudinal W bosons, for which the MSSM corrections are
largest, in Fig.\ \ref{fig:relscanGF} and differ only slightly from the
former result.

\section{Conclusions}
\label{sect:conclus}

The MSSM corrections to $\eeWW$ are probably too small to be detected at
LEP2, taking into account that the convolution with decay amplitudes in
the final and photon radiation in the initial state tends to smear the
corrections. 

With the projected accuracy of a linear collider, however, corrections of
this magnitude will play a role. Yet even a linear collider may have
a hard time disentangling the MSSM contributions because the one-loop
electroweak corrections include Sudakov logarithms of the form
$\log^2(s/m^2)$, whereas for SUSY corrections these logarithms are
power-suppressed \cite{CiC99}. Extrapolating from 20\% one-loop
electroweak corrections at TeV energies, the higher-order electroweak
contributions may be comparable in magnitude to the one-loop MSSM
corrections.

Although the calculation presented here
is limited to on-shell W bosons and hard-bremsstrahlung effects have not
been considered, these effects are the same for the SM and the MSSM, hence
it should be possible to use the MSSM cross-section presented here in a
straightforward way in existing Monte Carlo generators, for example
\cite{DeDRW00}. The Fortran code for this process is available at
{\tt http://www.hep-processes.de}.

\section*{Acknowledgements}

I thank W.~Hollik for discussions, C.~Schappacher for help with the
implementation of the MSSM parameter scan in Fortran, and G.~Weiglein for
proofreading the manuscript.

Parts of this calculation have been performed on the QCM computer cluster
at the University of Karlsruhe, supported by the Deutsche
Forschungsgemeinschaft (Forschergruppe ``Quantenfeldtheorie,
Computeralgebra und Monte-Carlo Simulation'').

\begin{flushleft}

\end{flushleft}

\end{document}